\documentstyle[12pt,aps,prb,epsf]{revtex}

\newcommand{\myfig}[3] {
\begin{figure}
\centerline{\epsfysize=#2  \epsfbox{#1.eps}}
\caption{#3}
\label{#1}
\end{figure} }

\def\be{ \begin{equation} }
\def\ee{ \end{equation} }

\begin{document}
\tightenlines
\draft

\title{\bf
Monte Carlo renormalization group study of
the Heisenberg and the XY antiferromagnet on the stacked triangular lattice
and the chiral $\phi^4$ model
}
\author{M.~Itakura}
\address{JSPS,
Center for Promotion of Computational Science and Engineering,
Japan Atomic Energy Research Institute,
Taito-ku, Higashiueno 6-9-3, Tokyo 110-0015, Japan\\}
\date{\today}
\maketitle

\begin{abstract}
With the help of 
the improved Monte Carlo renormalization-group scheme,
we numerically investigate
the renormalization group flow of 
the antiferromagnetic Heisenberg and XY spin model on the 
stacked triangular lattice (STA--model) and
its effective Hamiltonian,
$2N$--component chiral $\phi^4$ model which is used in the
field--theoretical studies.
We find that the XY-STA model with lattice size
$126 \times 144 \times 126$ exhibits clear first-order behavior.
We also find that the renormalization-group flow of the STA 
model is well reproduced by the chiral $\phi^4$ model, 
and that there are no
chiral fixed points of renormalization-group flow 
for $N=2$ and $3$ cases.
This result indicates that the Heisenberg-STA model also undergoes
first-order transition.

\end{abstract}
\pacs{PACS numbers: 75.10.Nr}

\section{Introduction}
The critical behavior of the antiferromagnetic 
%$N$-component
vector spin
models on the stacked triangular lattice (STA) 
is still a controversial
issue even after twenty years of extensive studies by
means of experimental, 
field-theoretical, 
and numerical methods. See 
\cite{tissier2000,pelissetto2001,plakhty2000,loison2000}
for recent works, and \cite{kawamurareview} for a review.

The field-theoretical 
renormalization group (RG) analysis tells
that when the number of spin component 
$N$ is greater than some threshold value $N_c$,
there are so-called "chiral fixed points" of the RG which
control the critical behavior
and are characterized by novel values of critical
exponents, while
for $N<N_c$
such  fixed points disappear and the phase transition
is of first order (see \cite{kawamurareview}for a review).
Theoretical estimations of $N_c$ have been made by various
authors by means of
$\epsilon$-expansion \cite{pelissetto2001,kawa-ft,antonenko2},
fixed dimensional perturbation \cite{pelissetto2001,antonenko1},
local potential approximation \cite{zumbach1993}, and
effective average action approach \cite{tissier2000},
but the estimated values range from
negative value to 6.5, depending on the method employed 
(however, recent results tend to be around 6.0).
Thus the critical behaviors of the physical relevant cases
$N=2$ (XY spin) and $N=3$ (Heisenberg spin) are still unclear.
A number of experimental studies \cite{plakhty2000}
and numerical simulations \cite{kawamuramc,plumer} have yielded
results which suggest second-order phase transition for $N=2$ and $3$.
However, recently it has been pointed out that
the critical exponent $\eta$, calculated 
from the scaling relation, becomes
negative in some of these results,
which is unphysical and indicates that
the observed critical behavior is in fact pseudo-critical
behavior induced by the slow RG flow 
\cite{tissier2000,loison2000,zumbach1993}.

The present work intends to clarify the issue by
numerically observing the RG 
flow and investigating whether there are chiral fixed points or not
for several values of $N$.
The paper is organized as follows: in the next section,  
we will present the method by which we observe the RG flow
in the Monte Carlo simulations.
Details of the Monte Carlo simulations are given in Sec.~\ref{mc}. 
In Sec.~\ref{results} the RG flow diagrams obtained from
the simulations
for $N=2,3, 8$ cases are presented.
The last section is devoted to the concluding remarks.

\section{Numerical observation of the RG flow}
\label{mcrg}
\subsection{chiral $\phi^4$ model}

The critical behavior of the $N$-component STA model is
essentially described by the following $2N$-component 
Ginzburg-Landau-Wilson Hamiltonian \cite{kawamurareview}:
\be
H=\int dx \left[
K \left(
(\nabla \vec{\phi}_a )^2
+
( \nabla \vec{\phi}_b )^2
\right)
+ r
\left( \vec{\phi}_a^2 + \vec{\phi}_b^2 \right)
+ u
\left( \vec{\phi}_a^2 + \vec{\phi}_b^2 \right)^2
+v
\left(
(\vec{\phi}_a \cdot \vec{\phi}_b)^2 -
\vec{\phi}_a^2 \vec{\phi}_b^2
\right)
\right]
,
\label{hamiltonian-c}
\ee
where $\vec{\phi}=(\vec{\phi}_a, \vec{\phi}_b)$ is $N+N$-component
vector field defined on the continuum space.
In the field-theoretical studies,
$K$ is fixed to unity to eliminate
ambiguity of coefficients induced by a trivial
rescaling of $\phi$,
namely, $\vec{\phi}\rightarrow c \vec{\phi}$.
In this regularization scheme, $r$ plays a role of temperature.
In the present work we concentrate on the case $v \geq 0$.
Figure \ref{rgth00} depicts
the renormalization group flow of the scaling variables
$r$, $u$, and $v$.
There are some trivial fixed points,
namely:
\begin{description}
\item[H]: high-temperature fixed point $(r, u, v)=(\infty, 0, 0),$
\item[L0]: anisotropic low-temperature fixed point $(-\infty,\infty,\infty),$
\item[L1]: isotropic low-temperature fixed point $(-\infty,\infty,0),$
\item[G]: Gaussian fixed point $(0,0,0).$
\end{description}
%They are denoted by
%H, L0, L1, and G, respectively.

The flow on the critical plane projected onto the $u$--$v$
plane is shown in Fig. \ref{rgth} for (a) $N<N_c$ and (b) $N>N_c$ cases.
When $N>N_c$,
there are stable and unstable chiral fixed points denoted by
$C_1$ and $C_2$, respectively,
beside the $O(2N)$ symmetric and the Gaussian fixed points
denoted by O and G, respectively.
The two fixed points $C_1$ and $C_2$ approach as $N$ decreases,
and annihilate each other at $N=N_c$.

In the isotropic $\phi^4$ models,
the correction-to-scaling term (distance to the
Wilson-Fisher fixed point) rapidly converges to zero
as $b^{-\omega}$, where $b$ is a renormalization factor
and $\omega \sim 0.8$ is the correction-to-scaling exponent.
Therefore the asymptotic critical behavior
can be observed in finite systems with moderate size.
However, when
an anisotropic quartic term is included in the Hamiltonian,
it in general induces very slow RG flow along the anisotropy
direction.
The correction-to-scaling exponent at an anisotropic fixed point
(if it exists) is usually of order $0.1$ or less,
and extremely large system is needed to observe
asymptotic critical behavior in the simulation of finite systems.
Even when there are no chiral fixed points
and the RG flow eventually diverges, the RG flow in the intermediate
region is very slow \cite{zumbach1993}
and it is very hard to observe asymptotic 
first-order behavior.
Thus, in the simulation of finite systems,
it is crucial to check the convergence to the fixed point
of the renormalized quartic coupling constants $u$ and $v$.
Now let us consider how to observe these renormalized coupling constants
in the numerical simulations.
In the continuum theory,
they describe the behavior of renormalized field variables
$\vec{\phi}({\bf x};l)$
defined as follows:
\be
\vec{\phi}({\bf x};l)\equiv
\int _{|{\bf k}|<1/l}
\exp(i {\bf k}\cdot {\bf x})
\vec{\phi}({\bf k}) d{\bf k},
\ee
where $l$ is a cut-off length
and $\vec{\phi}({\bf k})$ denotes Fourier
component of the field $\vec{\phi}$.
We denote by $u_l$ and $v_l$ the renormalized coupling constants
of cut-off length $l$: the renormalization flow 
in Fig. \ref{rgth} is a trajectory
of  $u_l$ and $v_l$ on the critical plane when $l$ is increased.
The renormalization behavior of these quantities
may be well observed via their conjugate quantities,
which are easier to observe in numerical simulations:
\be
\langle (\vec{\phi}^2({\bf x};l))^2 \rangle
,
\label{ulc-0}
\ee

\be
\langle
\vec{\phi}_a^2({\bf x};l) \vec{\phi}_b^2 ({\bf x};l)
 -(\vec{\phi}_a ({\bf x};l) \cdot \vec{\phi}_b ({\bf x};l))^2
\rangle
\label{vlc-0}
.
\ee
%%%%U,V
Of course, these quantities
are affected not only by $u_l$ and $v_l$ 
but also by other irrelevant scaling fields,
such as coefficients of higher order terms like $|\vec{\phi}|^6$ and
$|\vec{\phi}|^8$.
However, these irrelevant scaling fields rapidly vanishes 
(faster than the leading correction-to-scaling term in the
isotropic model, $\sim l^{-0.8}$) and their effects
can be ignored after sufficient renormalization
as long as $u_l$ and $v_l$ converge (or diverge) very slowly.

Note that the definition of the renormalized field
$\vec{\phi}({\bf x};l)$ needs scaling prefactor
so that the coefficients such as $u_l$ and $v_l$
remain finite when $l$ goes to infinity.
The following scaling scheme is simple and suitable for numerical
simulations \cite{prefactor}:
\be
\vec{\phi} ^\prime ({\bf x};l) \equiv
\frac{\vec{\phi}({\bf x};l)}
{\sqrt{
\langle
\vec{\phi}({\bf x};l)^2 \rangle
}}.
\ee
Substituting $\vec{\phi}({\bf x};l)$ by
$\vec{\phi} ^\prime ({\bf x};l)$,
equation (\ref{ulc-0}) and (\ref{vlc-0})
lead as follows:
\be
\frac{
\langle (\vec{\phi}^2({\bf x};l))^2 \rangle
}{
\langle (\vec{\phi}^2({\bf x};l)) \rangle ^2
}
,
\label{ulc}
\ee

\be
\frac{
\langle
\vec{\phi}_a^2({\bf x};l) \vec{\phi}_b^2 ({\bf x};l)
 -(\vec{\phi}_a ({\bf x};l) \cdot \vec{\phi}_b ({\bf x};l))^2
\rangle
}{
\langle (\vec{\phi}^2({\bf x};l)) \rangle ^2
}
\label{vlc}
.
\ee

Basically, in the present work,
we observe these quantities in the numerical simulations
of finite lattice version of the
Hamiltonian (\ref{hamiltonian-c}) defined as follows:

\be
H={K N \over 2} \sum_{<ij>} 
(\vec{\phi}(i)-\vec{\phi}(j))^2
+r\sum_i
\vec{\phi}(i)^2
+u \sum_i
\left(\vec{\phi}(i)^2 \right)^2
+v \sum_i
\left[
(\vec{\phi}_a(i) \cdot \vec{\phi}_b(i))^2 -
\vec{\phi}_a(i)^2 \vec{\phi}_b(i)^2
\right]
,
\label{hamiltonian-l}
\ee
where 
$\vec{\phi}(i)=\left(\vec{\phi}_a(i), \vec{\phi}_b(i) \right)$
is an
$N+N$-component vector defined on the lattice site i
of an $L\times L \times L$ cubic lattice with periodic boundary
condition being imposed,
the summation $\sum_{<ij>}$ runs over all nearest neighbor
pairs of the lattice site.
We use a restriction $u + v/4 -2r=0$ so that
the minimum of the 
Hamiltonian takes place at $|\vec{\phi}(i)|=1$ to eliminate
the ambiguity of the trivial rescaling of $\phi$.
Actually we use the following parameterization:
\be
r= -2\lambda ,\,\,\,\,
u= \lambda (1+A) ,\,\,\,\,
v =4 \lambda A
\ee
where $A$ controls the strength of the anisotropy
and $\lambda$ controls the "hardness" of the spin---
the larger $\lambda$ is, the smaller the fluctuation of
$|\vec{\phi}(i)|$ is.
A line $\lambda=0$ corresponds to $u-v/4=0$, on which instability
of the Hamiltonian (\ref{hamiltonian-l}) occurs; thus $\lambda$
should be positive.

In the lattice models, 
renormalized field corresponds to block spin variables.
For example, consider a block spin variable
$\vec{\phi}_{b,L}$
defined on 
a block of $b\times b\times b$ spins
in an $L\times L\times L$ lattice model.
One can 
calculate the quantities (\ref{ulc}) and (\ref{vlc})
using $\vec{\phi}_{b,L}$,
which we denote by $U_{b,L}$ and $V_{b,L}$, respectively,
and check the
convergence or divergence of the renormalization flow.
In the conventional MCRG scheme \cite{mcrg-conv},
one calculates (\ref{ulc}) and (\ref{vlc})
for fixed value of $L$ and increasing value of $b$.
Note that, after the ''renormalization'' by the block spin
transformation, the number of spins, $(L/b)^3$, decreases and
the Hamiltonian does not stay in the same Hamiltonian space,
unlike the continuum case.
Therefore the number of block spins must be sufficiently large,
i.e. $L/b \gg 1$, so that 
(\ref{ulc}) and (\ref{vlc})
 take the same value at the fixed point for different values of $b$.

The present author proposed another scheme  \cite{mcrg},
in which the ratio $L/b$ is fixed and 
both $L$ and $b$ are increased.
Let us denote a 
renormalized Hamiltonian which contains $l^3$ block spins
by $H_l(K,r,u,v, \{w_i\})$, where
$K,r,u,v$ are the renormalized coefficients
of terms which correspond to that in the original
Hamiltonian (\ref{hamiltonian-l}), and
$\{w_i\}$ denote set of coefficients of higher order irrelevant terms
such as $|\vec{\phi}|^6$, which is not included in the original Hamiltonian.
Now consider the behavior of
$\vec{\phi}_{L,L}$ and $\vec{\phi}_{2L,2L}$:
the behavior of both quantities are described by a
renormalized Hamiltonian which contains only 1 spin,
but with different coefficients.
Here,
$\vec{\phi}_{2L,2L}$ is a renormalized spin, obtained by 
applying real-space renormalization of  factor-$2L$ to the original
spins. Let us regard this factor-$2L$ renormalization as
two successive renormalizations, namely factor-$2$ renormalization
at first and then factor-$L$ renormalization.
Accordingly, the original Hamiltonian $H_{2L}(K,r,u,v, \{0\})$
is at first renormalized to
$H_L(K^\prime,r^\prime,u^\prime,v^\prime, \{w_i\})$,
then again renormalized to
\def \pp {{\prime \prime}}
$H_{1}(K^\pp, r^\pp,u^\pp,v^\pp, \{w_i ^\prime\})$.
If the starting Hamiltonian is at a fixed point
and $L$ is sufficiently large,
coefficients before and after the factor-$2$ renormalization
should remain unchanged
except for the higher order coefficients $\{w_i ^\prime\}$ \cite{note1}.
At this stage, $U_{2L,2L}$ and $V_{2L,2L}$ can be calculated  
(in other words, factor-$L$ renormalization)
from $H_L(K^\prime,r^\prime,u^\prime,v^\prime, \{w_i\})$,
while $U_{L,L}$ and $V_{L,L}$ are calculated
from $H_L(K,r,u,v, \{ 0 \})$.
Since the coefficients of these two Hamiltonians are the same
(except the irrelevant ones), $U_{2L,2L}=U_{L,L} + O(L^{-\omega})$ and
$V_{2L,2L}=V_{L,L} + O(L^{-\omega})$ should be satisfied, where
$\omega$ denotes correction-to-scaling exponent of the 
higher order terms such as $|\vec{\phi}|^6$.
The so-called "phenomenological renormalization group method"
\cite{binder} uses this property to locate the critical point.

When the Hamiltonian is not at a fixed point,
the coefficients 
before and after the first factor-$2$ renormalization
differ: they moves along the renormalization flow.
This difference is reflected by $U_{L,L}$ and $V_{L,L}$;
If the renormalization flow slowly converges to some fixed point,
$U_{L,L}$ and $V_{L,L}$ slowly converge to some fixed value,
while when the phase transition is of first-order and the
renormalization flow diverges, so do $U_{L,L}$ and $V_{L,L}$ \cite{q4ita}. 
We will henceforth denote these quantities simply by  $U_L$ and $V_L$,
and investigate their behavior at the critical point
when $L$ is increased.
Here their explicit form is written down for the readers' convenience:
\be
U_L={
\langle \left(\vec{M}^2\right)^2 \rangle
\over
\langle \vec{M}^2 \rangle ^2
}
\label{ul},
\ee
\be
V_L={
\langle
\vec{M}_a^2 \vec{M}_b^2 -(\vec{M}_a \cdot \vec{M}_b)^2
\rangle
\over
\langle \vec{M}^2 \rangle ^2
}
\label{vl},
\ee
where $\vec{M}=(\vec{M}_a, \vec{M}_b)=\sum_i \vec{\phi}(i)$.
In the past numerical studies
in which only a specific model was investigated,
the only way to reach the final fixed point (if it exists)
or to observe asymptotic first-order behavior was to simulate
extremely large systems, and it was impossible in general.
In the present work, we investigate various Hamiltonians
and scan the Hamiltonian space in search of fixed points.
This method allows us to determine the asymptotic critical behavior
by simulations of moderately large systems.

Note that,
in the theoretical works
one can concentrate on the RG flow on the critical plane
on which inverse susceptibility vanishes,
while in the numerical works one must determine the critical
point from the numerical data.
For this purpose we use a quantity which corresponds to the 
following quantity in the continuum theory:
\be
\frac{
\langle
\int _{|k|<1/l} ({\bf k}/l)^2 \vec{\phi}^2 ({\bf k})
d^D{\bf k}
\rangle
}{
\langle (\vec{\phi}^2({\bf x};l)) \rangle
}
=
l^2
\frac{
\langle
(\nabla \vec{\phi} ({\bf x};l)) ^2
\rangle
}{
\langle (\vec{\phi}^2({\bf x};l)) \rangle
}
.
\ee

In the lattice models, the above quantity corresponds to
the normalized correlation between two adjacent block spins.
Actually we observe the following quantity:
\be
C_L=
{
\langle
\vec{\phi}({\bf k}_1)\cdot \vec{\phi}(-{\bf k}_1)
\rangle
\over
\langle \vec{M}^2 \rangle
}
\label{cl},
\ee
where 
$\vec{\phi}({\bf k_1}) = \sum_{\bf r} \vec{\phi}(r)
\exp (i {\bf k_1}\cdot {\bf r})$
with ${\bf k}_1 = (2\pi/L,0,0)$.

Figure \ref{rgmc0} depicts
the RG flow in $U_L$--$V_L$--$C_L$ space:
a number of trivial fixed points, namely,
high-temperature fixed point 
$(1+1/2N,{(2N+1)(N-1)\over 8N(N+1)},1)$,
isotropic low-temperature fixed point $(1,{N-1 \over 4(N+1)},0)$,
and anisotropic low-temperature fixed point
$(1,{1 \over 4},0)$ are denoted by H, $L_0$, and $L_1$, respectively.
Beside these points, there are some trivial manifolds:
when the probability distribution of the order parameter $\vec{M}$
is isotropic, a relation $V_L={N-1 \over 4(N+1)}U_L$ is satisfied,
while in the strong anisotropy limit where
$\vec{M}_a \cdot \vec{M}_b=0$ and $|\vec{M}_a|=|\vec{M}_b|$ hold,
$V_L=U_L/4$ is satisfied.
At the Gaussian fixed point,
the behavior of the finite system is governed by the zero-mode \cite{mcrg},
therefore the Gaussian fixed point lies on the $C_L=0$ plane.
The RG flow on the critical plane, projected onto the
$U_L$--$V_L$ plane, is shown in Fig. \ref{rgmc}
for (a) $N<N_c$ and (b) $N>N_c$ cases.
In both cases, 
the RG flow along the (approximately) horizontal
direction rapidly converges, while
along the (approximately) vertical direction
the flow is expected to slowly converge/diverge.

To obtain the RG flow on the critical plane, 
we investigate the Hamiltonian (\ref{hamiltonian-l})
for various values of $A$ and $\lambda$,
tuning $K$ so that $C_{L}=C_{2L}$ is satisfied
and observe the difference between $(U_L, V_L)$  
and $(U_{2L}, V_{2L})$.
It should be noted that the above definition of 
critical plane induces systematic 
deviation. For example, consider the RG flow of 
the isotropic $\phi^4$ model (set $v=0$ in (\ref{hamiltonian-l}) ).
The flow of $(U_L, C_L)$ near the Wilson-Fisher fixed point
is depicted in Fig. \ref{rg-phi4} (a) \cite{mcrg}.
If one define the critical point by $C_{L}(K)=C_{2L}(K)$,
the flow of $U_L$ becomes the one shown in Fig. \ref{rg-phi4}(b),
thus the arrow tends to ``overshoot'' the RG fixed point.
However, the direction of the flow can be correctly estimated, and
the systematic error vanishes as one approaches the fixed point.

\subsection{STA models}

We also observe the RG flow of the
following STA model:
\be
H=-K \sum_{<ij>} J_{ij}\vec{S_i}\cdot \vec{S_j} ,
\ee
where 
$\vec{S_i}$ denotes a two-component (XY) or
three-component (Heisenberg) vector spin 
with $|\vec{S}_i|=1$ defined on
the lattice site $i$ of the stacked triangular lattice.
We set $J_{ij}=-1$ (antiferromagnetic)
for intra-plane nearest neighbor pairs,
$J_{ij}=3/4$ (ferromagnetic)
for inter-plane nearest neighbor pairs, and $J_{ij}=0$ otherwise,
so that the Fourier
transform of $J_{ij}$ near the antiferromagnetic mode \cite{kaf}
$\vec{k}_{AF}=(2\pi/3,0,0)$ becomes isotropic in the $\vec{k}$-space, i.e.
$J(\vec{k}_{AF}+\vec{k})\sim J(\vec{k}_{AF})+ c|\vec{k}|^2 +O(|\vec{k}|^4)$.
Note that the critical values of 
quantities such as $U_L, V_L,$ and $C_L$ 
do not depend on the microscopic lattice structure,
but {\it do} depend on the macroscopic lattice structure
such as boundary conditions and aspect ratios \cite{ko01,kok99}.
Therefore we use rectangular system 
which contains $L_x \times L_y \times L_z$ spins
(see Fig. \ref{tri8}), imposing periodic boundary condition
for all three directions. The aspect ratio then becomes
$L_x:\sqrt{3}L_y/2:L_z$. We simulate the system
with $(L_x,L_y,L_z)=(21,24,21)$, $(42,48,42)$, $(84,96,84)$, 
and $(126,144,126)$,
all of which give aspect ratio $1:0.99\cdots :1$.

The order parameter is defined as follows \cite{kawamurareview}:
\be
\vec{M}_a= \mbox{Re}[\vec{S}(\vec{k}_{AF})]=
\vec{M}_A - {1 \over 2} \vec{M}_B -{1\over 2} \vec{M}_C,
\ee
\be
\vec{M}_b= \mbox{Im}[\vec{S}(\vec{k}_{AF})]=
 {\sqrt{3}\over 2} \vec{M}_B -{\sqrt{3}\over 2} \vec{M}_C,
\ee
where 
$\vec{S}(\vec{k}_{AF})$ denotes Fourier component of $\vec{S}_i$
at $\vec{k}_{AF}$,
and $\vec{M}_A,\vec{M}_B,\vec{M}_C$ denote the magnetization
on the three sublattices. Then $U_L$ and $V_L$ are calculated from
Eq. (\ref{ul}) and Eq. (\ref{vl}).
$C_L$ is calculated as follows:
\be
C_L={
\langle
|\vec{S}(\vec{k}_{AF}+\vec{k}_1)|^2+
|\vec{S}(\vec{k}_{AF}-\vec{k}_1)|^2
\rangle
\over \langle \vec{S}^2(0)\rangle
},
\ee
where $\vec{k}_1$ is the smallest, non-zero momentum in the $\vec{k}$-space.
We observed the values of $C_L$ for three kinds of direction, namely,
$\vec{k}_1=({2\pi \over L_x},0, 0)$,
$(0, {4\pi \over \sqrt{3} L_y},0)$, and
$(0, 0, {2\pi\over L_z})$, and confirmed that they
coincide each other within statistical errors.
This indicates that the simulated system is spatially isotropic. 

%MMMM
\section{Monte Carlo simulation}
\label{mc}
\subsection{chiral $\phi^4$ models}
We used the usual single-spin update Metropolis algorithm,
since there is no spin reflection axis which preserves the 
anisotropy term in (\ref{hamiltonian-l}) and
the cluster algorithm \cite{wolff} cannot be applied.
In a single-spin update process, the new spin value was chosen as
follows:
\be
\vec{\phi}_{new}(i)= \vec{\phi}_{old}(i)+ \frac{R_g}{\sqrt{2N}} \vec{G}
\ee
where $\vec{G}$ is an $2N$-component vector whose components
are independent Gaussian random variables with average $0$
and variance $1$.
The optimum value of the amplitude of random move, $R_g$,
was determined by simulating small systems.
% to minimize the autocorrelation.
The optimum value was found to be 
$R_g \approx 0.3$ for all $N$, $A,$ and $\lambda$.
Each Metropolis sweep was followed by one overrelaxation-type
update \cite{caselle}. Hereafter we refer
one Metropolis sweep plus one overrelaxation-type sweep as ``MCS''.

Physical quantities were observed
for every $C_{INT}\times L^2$ MCS,
where $C_{INT}$ is a constant 
which is adjusted so that
the correlations between successively observed quantities 
become less than $0.70$, namely:
\be
\frac{ 
\langle X_t X_{t+1} \rangle
-\langle X_t \rangle ^2
}{
\langle X_t^2 \rangle
-\langle X_t \rangle ^2
}<0.7, \label{ac}
\ee
where $X_t$ denotes observed value of a 
physical quantity $X$ at the $t$th observation.
The value of $C_{INT}$ varies from $0.1$ to $2.0$, depending mainly on $A$. 
Actual values will be summarized in the later sections.
For each values of $L$, $N$, $A$, and $\lambda$,
physical quantities were observed 
$10^4$ to $10^5$ times, depending on the
accuracy required to observe the RG flow.
Statistical errors were estimated by the Jackknife method.
Histogram method \cite{hist} was used to calculate thermal averages
at $K$ slightly away from that where the simulation is actually
carried out.

Simulations were mainly carried out on 
Fujitsu VPP5000 vector processors at JAERI.
The checkerboard-wise decomposition of the lattice
was used for vectorization.
Furthermore, we simultaneously 
simulated a number of independent systems,
whose spins were stored into one long vector.
This promoted the vectorization, especially
when $L$ is small, and accelerated the calculations.
For $N=8$ and $L=16$ system, one MCS
took 0.84 ms on the VPP5000.

\subsection{STA models}
In the frustrated spin models,
the cluster update algorithm tends to flip all the spins
and does not accelerate the simulation.
Therefore we again used only the single-spin update Metropolis algorithm,
followed by an
overrelaxation sweep, i.e. 180 degree rotation of a spin $\vec{S}_i$
with respect to
its local field $\sum_j J_{ij} \vec{S}_j$.
In the Metropolis update, the new spin direction was chosen as follows:
\be
S_x = \sqrt{1-z^2} \cos(\theta),  \,\,\,
S_y = \sqrt{1-z^2} \sin(\theta),  \,\,\,
S_z = z,
\ee
for Heisenberg case and
\be
S_x =  \cos(\theta),  \,\,\,
S_y =  \sin(\theta)  , 
\ee
for XY case,
where $z$ and $\theta$ are uniformly distributed random numbers
in the range $[-1,1)$ and $[0, 2\pi)$, respectively.
Physical quantities were observed for every $L_x \times L_y /20$
MCS for the Heisenberg case and
$L_x \times L_y /10$ MCS for the XY case.
This interval was long enough to satisfy
the condition (\ref{ac}) as long as 
the energy histogram has no double-peak.
Physical quantities were observed $10^4$ times
for all system sizes up to $(L_x,L_y,L_z)=(42,48,42)$.

To observe the energy histogram of the larger system,
we also simulated very large systems $(L_x,L_y,L_z)=(84,96,84)$
 and $(126,144,126)$ for the XY case.
Owing to the prominent double-peak of the energy histogram,
autocorrelation time was much longer than $L_x \times L_y /10$ MCS
in these sizes.
We used $4\times 10^6$ MCS for the $(84,96,84)$ case
and $7 \times 10^6$ MCS for the $(126,144,126)$ case,
which were
1000 times longer than the auto-correlation time.
One MCS of $N=3$ and $(L_x,L_y,L_z)=(84,96,84)$ system
took 27 ms on the VPP5000.

\subsection{Stiefel's $V_{N,2}$ models}
Stiefel's $V_{N,2}$ model \cite{loison2000}
corresponds to the Hamiltonian (\ref{hamiltonian-l})
with $\lambda=\infty$ and $A=\infty$, in which 
restrictions $\phi_a \cdot \phi_b=0$,
$\phi_a^2=\phi_b^2=1/2$ are imposed to all the spins.
We also simulated this model and investigated the RG flow.
In the Monte Carlo simulation, only the single-spin
Metropolis update was used.
We only simulated the $V_{3,2}$ model,
since the $V_{2,2}$ model is known to exhibit strong first-order
behavior \cite{loison2000}.
In the Metropolis update,
the new spin value was chosen as follows:
\begin{eqnarray}
\vec{\phi}_a&=&
( z, \sqrt{1-z^2} \cos\theta_1,\sqrt{1-z^2} \sin\theta_1)/\sqrt{2},\\
\vec{\phi}_b&=&
( \sqrt{1-z^2} \sin\theta_2,
 -\sin\theta_1\cos\theta_2-z \cos\theta_1\sin\theta_2,
\cos\theta_1\cos\theta_2-z\sin\theta_1\sin\theta_2)/\sqrt{2}
\end{eqnarray}
where $z$ and $\theta_{1,2}$ are uniformly distributed random numbers
in the range $[-1,1)$ and $[0, 2\pi)$, respectively.
Physical quantities were observed for every $L^2$
Metropolis sweeps. This interval
was long enough to
satisfy the condition (\ref{ac}) for all system sizes
up to $L=48$.
Physical quantities were observed $10^4$ times for these system sizes.

To observe the energy histogram of the larger system,
we also simulated a very large system $L=64$ and $L=80$.
Like the STA-XY case,
a prominent double-peak of the energy histogram emerges and
autocorrelation time gets much longer than $L^2$ MCS in these sizes. 
We used $6\times 10^6$ MCS for the $L=64$ case and
$2 \times 10^7$ MCS  for the $L=80$ case,
which were about 1000 times longer than the auto-correlation time.
For the largest size $L=80$, one Metropolis sweep took 22 ms on the VPP5000.

\section{results}
\label{results}
\subsection{XY case}
Parameters $\lambda$, $A$, and $K$ at which simulations were
carried out are summarized in Table \ref{n2param}.
Figure \ref{xyhist} shows the histogram of 
energy per spin, $E=-\sum J_{ij}\vec{S}_i \vec{S}_j/L_x L_y L_z$
of the STA-XY model with
$(L_x,L_y,L_z)=(126,144,126)$
and $(84,96,84)$ at $K=0.77262$.
It can be seen that the valley between the two peaks
deepens as the system size increases.
This is a clear evidence of first-order transition.
This peak does not appear when the size is smaller,
therefore the double peak has not been observed in the
past studies of STA-XY model in which smaller systems have been used.
Note that the first-order behavior has already been observed
in the Monte Carlo simulations of other models
which possess the same symmetry as the STA-XY model,
such as Stiefel model and restricted STA model \cite{loisonxy},
and quasi-one dimensional STA model \cite{plumer1d}.
These models
have stronger anisotropy than the STA-XY model
and the first-order behavior can be observed in smaller systems.

The two peaks of the histogram correspond to
ordered and disordered phases, and the energy difference between
the peaks provides lower bound of the latent heat per spin $\Delta$. 
On the other hand, experimental measurement of specific heat
of the STA-XY materials such as CsMnBr$_3$ or Holmium \cite{wang}
indicates second-order transition or very weak
first-order transition whose latent heat is beneath the resolution
of measurements, thus the upper bound of the latent heat is given.
To compare the strength of the first-order transition,
an adimensional parameter $c=\Delta/k_B T_c$ is usually used,
where $k_B$ denotes Boltzmann constant and $T_c$ denotes the
transition temperature.
From Fig. \ref{xyhist}, the upper bound of the latent heat per spin is
estimated as $\Delta > 7.8 \times 10^{-3}$ and $c > 6.0 \times 10^{-3}$,
while Ref.\cite{wang} provides estimates $c < 2.9 \times 10^{-5} $
for CsMnBr$_3$ and  $c < 2.8 \times 10^{-5} $ for Holmium.
These {\it upper} bound of $c$ obtained from experiments is far smaller
than the {\it lower} bound of $c$ obtained in the present work,
thus there is an  apparent contradiction between them.

Note that the value of $c$ may depends on the strength of
the easy-plane anisotropy of the spin and the ratio between
inter and intra layer coupling constants.
The result of Ref. \cite{plumer1d} indicates that
when the inter-layer coupling is much larger than the
intra-layer coupling, the first-order transition becomes
stronger.  On the other hand, a weak easy-plane anisotropy
will lead to Heisenberg-to-XY crossover which weaken the transition,
since the transition is weakened as the number of spin component
is increased and eventually becomes second-order. Actually, we will
see in the following section
that the STA-Heisenberg model does not exhibit as strong a transition 
as the STA-XY model.
Thus the most plausible explanation of the apparent discrepancy
between the experimental results and the present work  may be that
the finiteness of the easy-plane anisotropy in the experimental
materials induces Heisenberg-to-XY crossover and the transition
is weakened.

Figure \ref{n2mcrg-s} and \ref{n2mcrg-w} 
show the RG flow of $(U_L, V_L)$ 
at the strong and weak anisotropy region, respectively, 
of the $N=2$ chiral $\phi^4$ model.
Figure \ref{n2mcrg-s} also shows the same plot for the STA-XY model.
One can see that the STA-XY model is on the ``runaway''
trajectory, which is another evidence of first-order transition.
It can also be seen that
there are no stable fixed points other than the
$O(4)$ symmetric one.
This means that any models or materials which possess
the same symmetry 
as STA-XY model, namely $Z_3 \times O(2)$, 
exhibit first-order transition.

Fig. \ref{n2vv} shows the vertical velocity of the RG flow
$V_{2L}-V_L$ plotted against $V_L$. 
The plots for the two different sizes $L=6$ and $L=8$
coincide within statistical errors.
This indicates that the system size is large enough to eliminate
finite-size artifact of the RG flow.
Thus we can safely extrapolate the asymptotic critical behavior
from the finite-size RG flow.
Moreover, the plots of the chiral $\phi^4$ model and the STA-XY
model are essentially the same. This means that the critical behavior
of the STA-XY model is well reproduced by the chiral $\phi^4$ model.

\subsection{Heisenberg case}
Parameters $\lambda$, $A$, and $K$ at which simulations were
carried out are summarized in Table \ref{n3param}.
Unlike the STA-XY case,
the STA-Heisenberg model 
does not show the double-peak behavior within the simulated
size $(L_x,L_y,L_z)=(84,96,84)$.
However, the $V_{3,2}$ model with $L=80$  shows a clear
double-peak behavior. Figure \ref{v32hist} shows the histogram of
energy per spin, $E=3 \cdot L^{-3}\sum J_{ij} (\vec{S}_i -\vec{S}_j)^2$
at $K=0.217685$. Although it is necessary to
simulate larger system and show that the 
valley between the two peaks deepens
to confirm the first-order transition,
this double-peak strongly indicate that
the transition is of first-order.

Figure \ref{n3mcrg-s} and \ref{n3mcrg-w} show the RG flow of
$(U_L, V_L)$
at the strong and weak anisotropy region, respectively,
of the $N=3$ chiral $\phi^4$ model.
Figure \ref{n3mcrg-s} also shows the same plot for
the STA-Heisenberg model and the $V_{3,2}$ model.
It can be seen that
the STA-Heisenberg model and the $V_{3,2}$ model
are on the ``runaway''
trajectory, therefore undergo first-order transition in the 
thermodynamic limit.
It can also be seen that
there are no fixed points other than the
$O(6)$ symmetric one.
This means that any models or materials which possess
the same symmetry 
as STA-Heisenberg model, namely $Z_3 \times O(3)$, 
exhibit first-order transition.
From Fig. \ref{n3mcrg-s}, we roughly
estimate that $L=96$ STA-Heisenberg
model needs factor-2 renormalizations 
three or more times to reach
the strong first-order region, which means that
the clear first-order behavior will be observed
in an $L \sim 800$ system.

Fig. \ref{n3vv} shows the vertical velocity of the RG flow
$V_{2L}-V_L$ plotted against $V_L$.
In this case, this quantity depends not only on $V_L$ but also
on $U_L$, although the dependence is not so strong. Therefore the plot
can not be fitted to a single line well.
However, there is no systematic deviation
between the plots of the two different sizes.
Thus we can safely extrapolate the asymptotic critical behavior
from the finite-size RG flow.
As in the XY case,
the plots of the chiral $\phi^4$ model and the STA-Heisenberg
model are essentially the same and the critical behavior
of the STA-Heisenberg model is well reproduced by the chiral $\phi^4$ model.

\subsection{$N=8$ case}
It is necessary to demonstrate that
the MCRG scheme presented in Sec. \ref{mcrg}
is able to detect the 
chiral fixed point if it does exists.
For this purpose, we investigate $N=8$ case,
which is greater than any theoretical estimate of $N_c$
in literature
\cite{tissier2000,pelissetto2001,kawa-ft,antonenko2,antonenko1,zumbach1993}.
Parameters $\lambda$, $A$, and $K$ at which simulations were
carried out are summarized in Table \ref{n8param}.
Figure \ref{n8mcrg} shows the RG flow of 
$(U_L, V_L)$ for $L=8 \rightarrow 16$.
The stable chiral fixed point $C_1$
and the unstable chiral fixed point $C_2$
are clearly seen.

Fig. \ref{n8vv} shows the vertical velocity of the RG flow
$V_{2L}-V_L$ plotted against $V_L$, using only the data
near the stable fixed point $C_1$.
Unlike the $N=2$ and the $N=3$ case, this plot has a zero.
The plots of the two different size
coincide within statistical error, which indicates that
the simulated system is large enough to observe the RG flow.

%%%%%CONC
\section{conclusion}
The large scale 
Monte Carlo simulation of the STA-XY
model has revealed that
the transition is of first-order.
This is the first numerical work to confirm the
first-order transition of the simple and isotropic STA-XY model.
%XXX
The result also indicate that the transition is 
strong enough to be observed in experiments.
But the finiteness of the easy-plane spin anisotropy of the
actual materials will weaken the transition compared to the
STA-XY model simulated in the present work,
in which the easy-plane anisotropy is infinitely strong.

We have also investigated the RG flow the 
$2N$--component chiral $\phi^4$ model
numerically and have found that
there are no chiral fixed points in $N=2$ and $N=3$ cases,
and we conclude that the STA-Heisenberg model also undergoes
first-order transition.
By a rough estimate, we expect that the STA-Heisenberg model
exhibits strong first-order behavior when the size is larger than
$L=800$.
For $N=8$ case, we have found the chiral fixed points,
thus the value of $N_c$ is estimated as $3 <N_c < 8$.
Further study to determine the precise value of $N_c$ may be
of some interest, because such a study will serve to
test the accuracy of field theoretical techniques.
However, once the nature of physical relevant case $N=3$ and $N=2$
is clarified, the precise value of $N_c$ is 
of less importance.

\section {Acknowledgments}
The present author is 
grateful to 
D.~Loison, B.~Delamotte, K.~B.~Varnashev, and
H.~Kawamura
for useful comments.

%%%%%%%%%%%%%%%%%%%%%%%%%%%%%%%%%%%
\newpage

\myfig{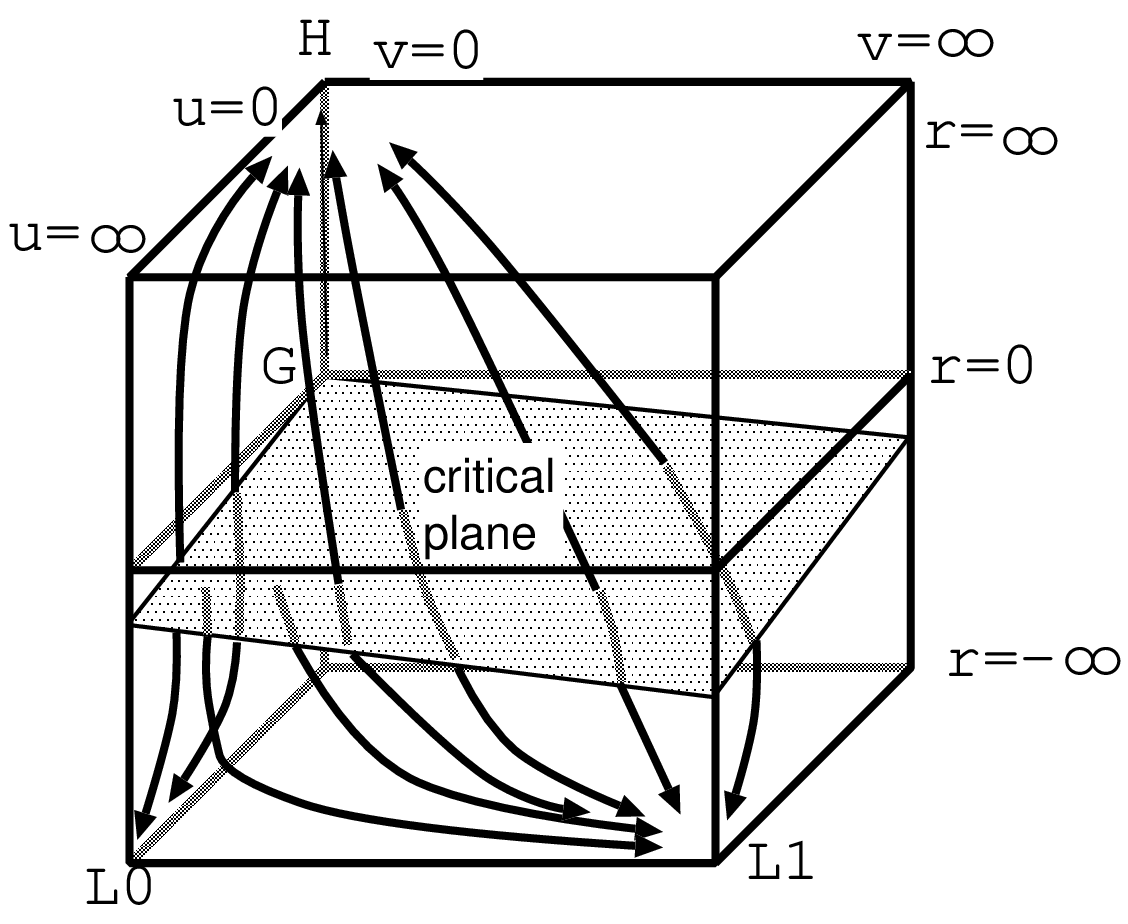}{6cm}{Renormalization flow
of $r$, $u$, and $v$.
H, G, L$_0$, and L$_1$
denote the high-temperature, Gaussian,
isotropic low-temperature, and anisotropic low-temperature
fixed point, respectively.
}

\myfig{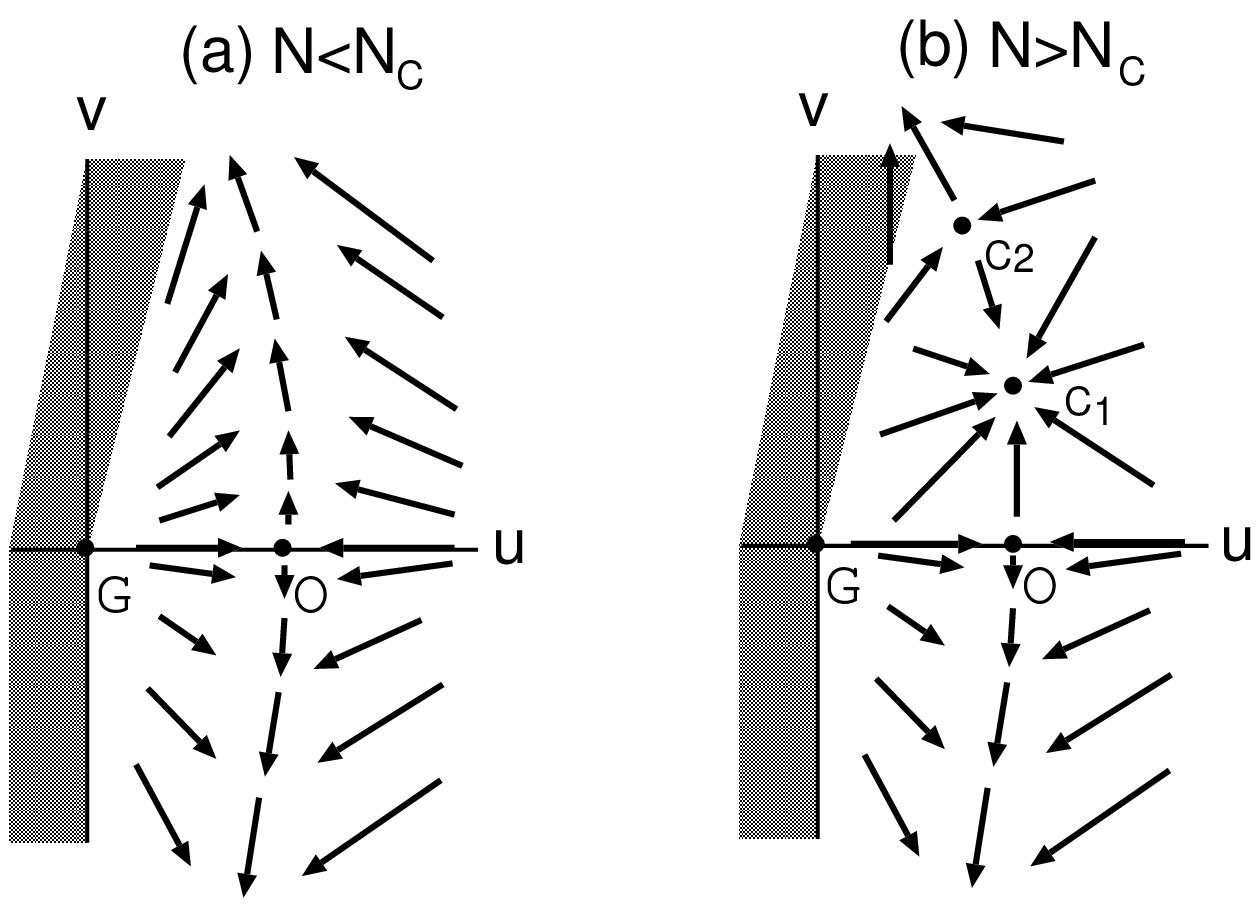}{6cm}{Renormalization flow
on the critical plane
projected onto the $u$--$v$ plane
 for (a) $N<N_c$ and (b) $N>N_c$.
Shaded area is unstable region.
G, O, $C_1$, and $C_2$ denote
Gaussian, $O(2N)$-isotropic, stable chiral,
and unstable chiral fixed points, respectively.
}

\myfig{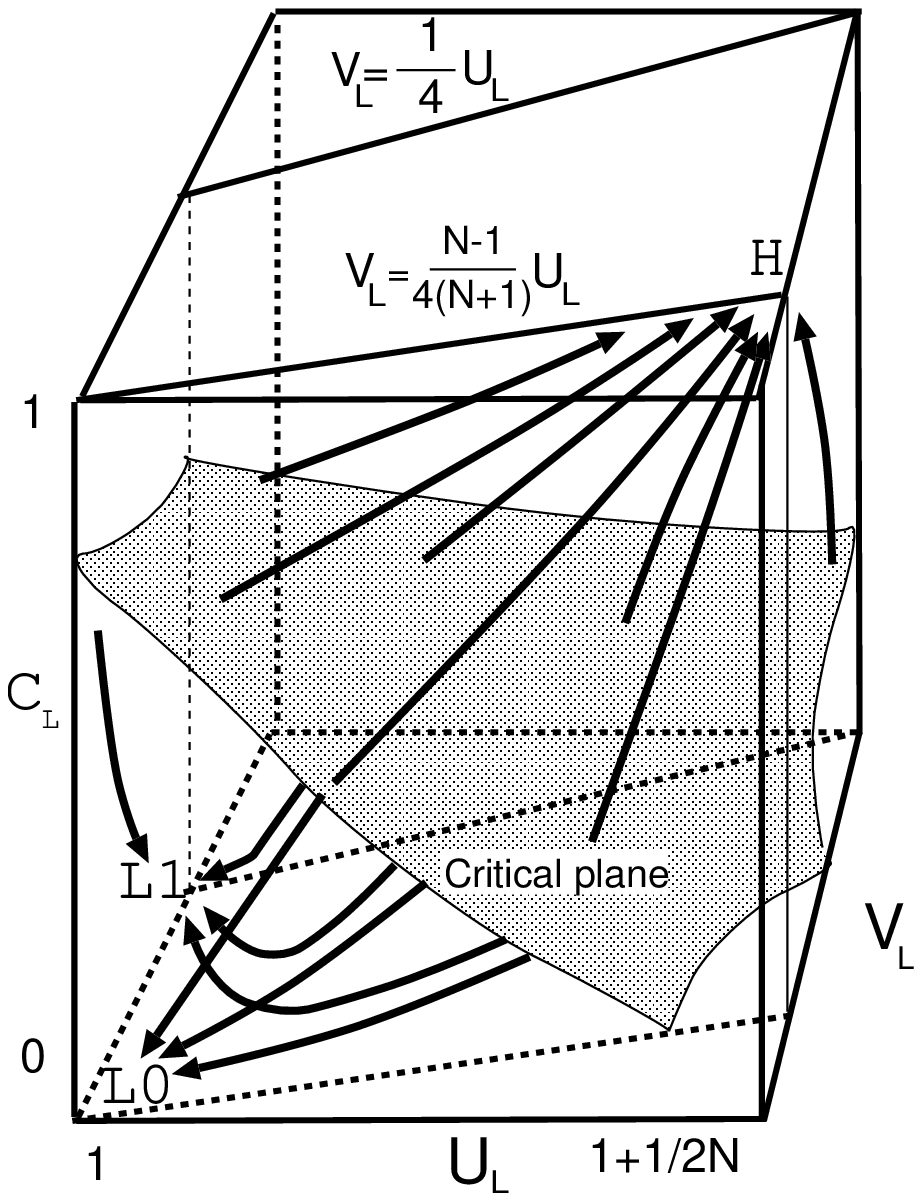}{6cm}{Renormalization flow
of $C_L$, $U_L$, and $V_L$.
}

\myfig{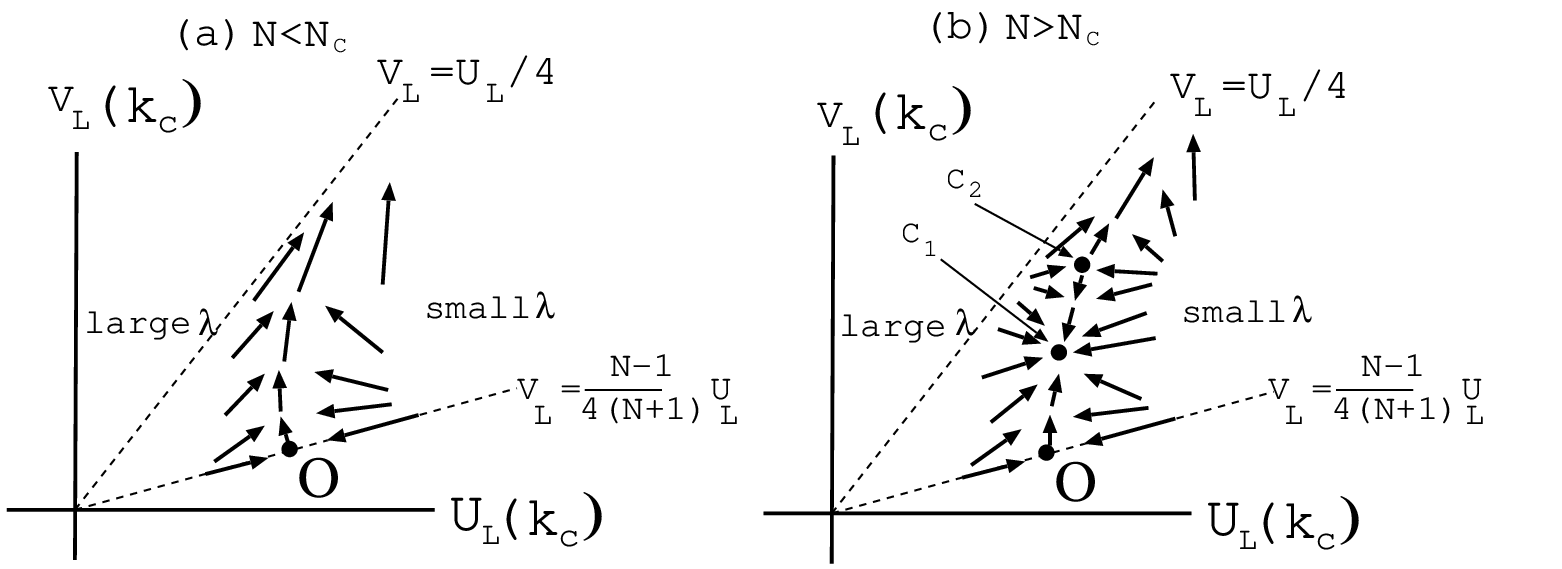}{6cm}{Renormalization flow
on the critical plane
projected onto the $U_L$--$V_L$ plane 
 for (a) $N<N_c$ and (b) $N>N_c$ cases.
The symbol O, $C_1$, and $C_2$ denote
 $O(2N)$-isotropic, stable chiral,
and unstable chiral fixed point, respectively.
}

\myfig{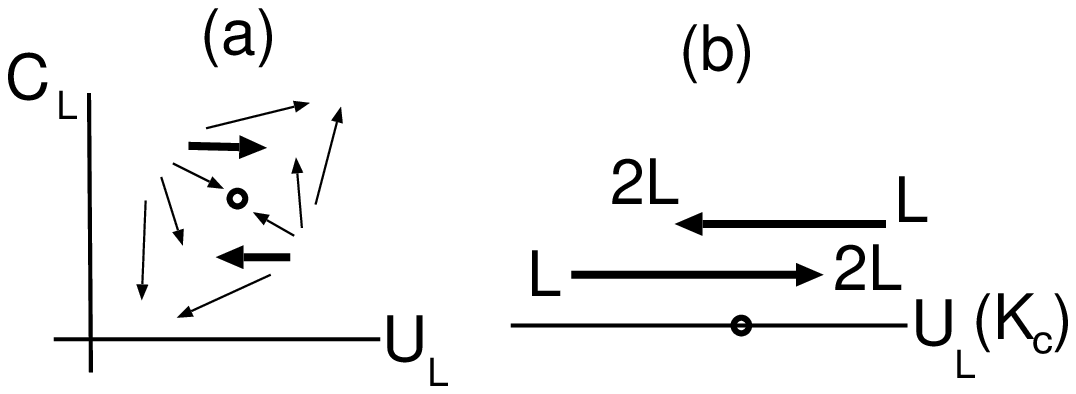}{6cm}{
(a) Renormalization flow
of $C_L$ and $U_L$
for isotropic model $v=0$ near the Wilson-Fisher fixed point.
(b) RG flow of $U_L(K_c)$ tends to ``overshoot''
the fixed point (see the main text).
}

\myfig{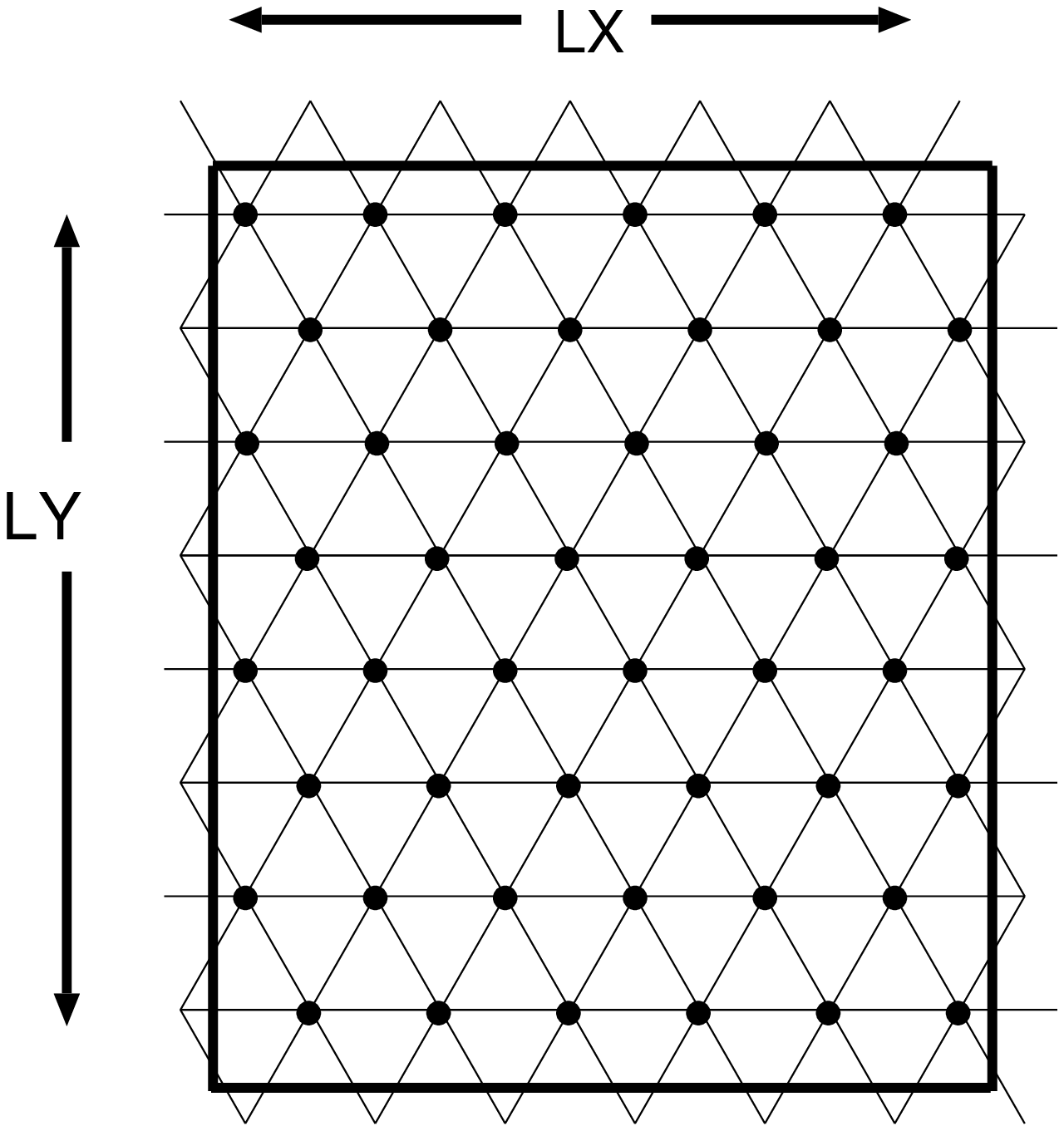}{6cm}
{
Rectangular system used for the
simulation of stacked triangular antiferromagnet.
The figure shows a system with $L_x=6, L_y=8$.
This layer is stacked $L_z$ times.
}

\myfig{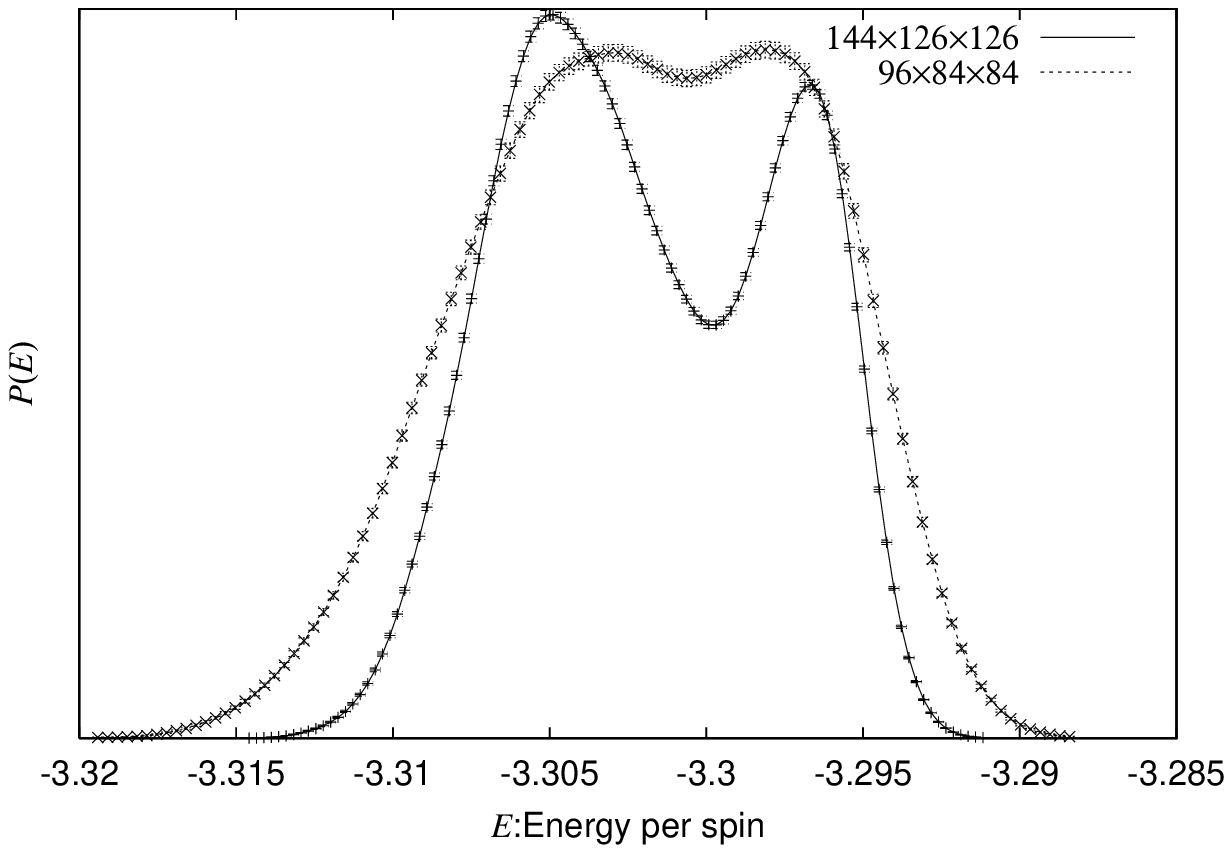}{12cm}
{
Histogram of
energy per spin of the STA-XY model with the size 
$(L_x,L_y,L_z)=(84,96,84)$ and $(126,144,126)$ at $K=0.77262$.
}
\newpage

\myfig{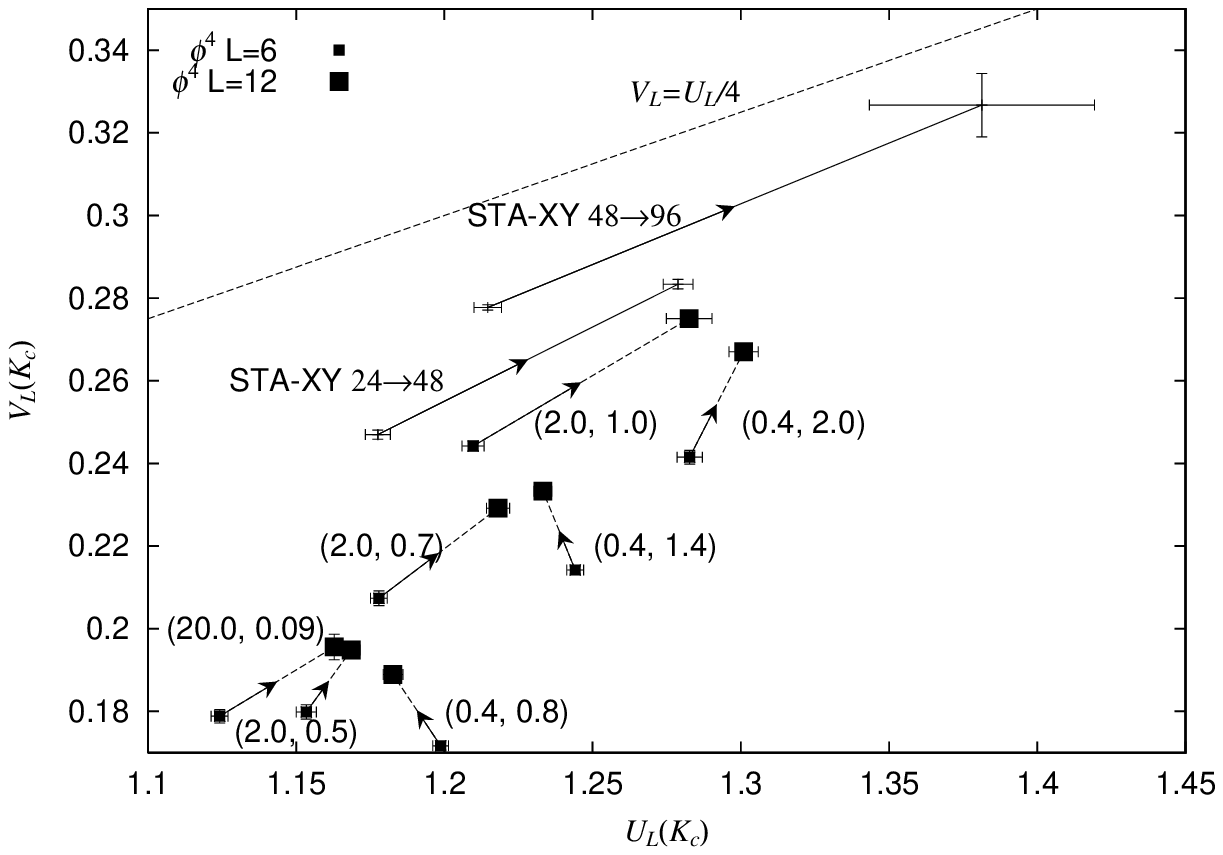}{10cm}
{
RG flow of the $N=2$ chiral $\phi^4$ model
and the STA-XY model
at the strong anisotropy region.
The numbers in parenthesis are the values of $(\lambda, A)$
of the chiral $\phi^4$ model.
}

\myfig{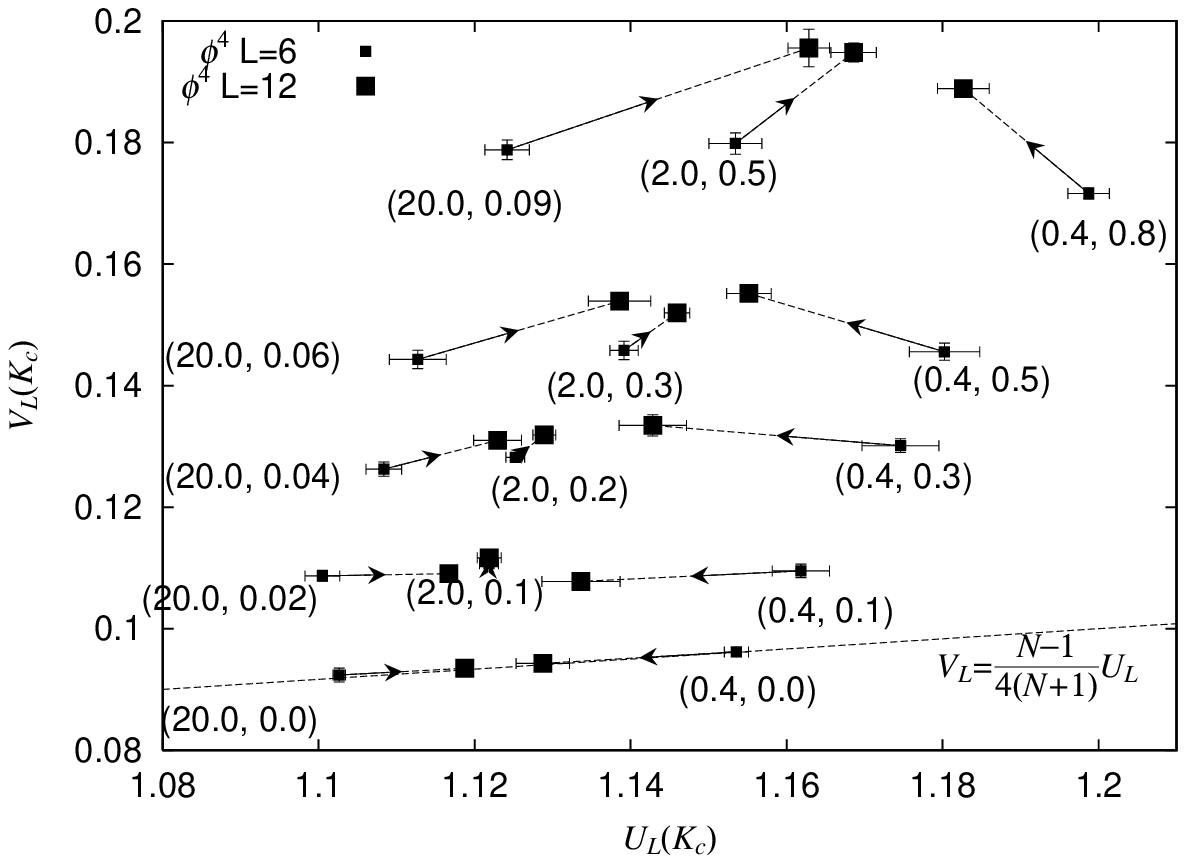}{10cm}
{
RG flow of 
the $N=2$ chiral $\phi^4$ model
at the weak anisotropy region.
The numbers in parenthesis are the values of $(\lambda, A)$.
}

\myfig{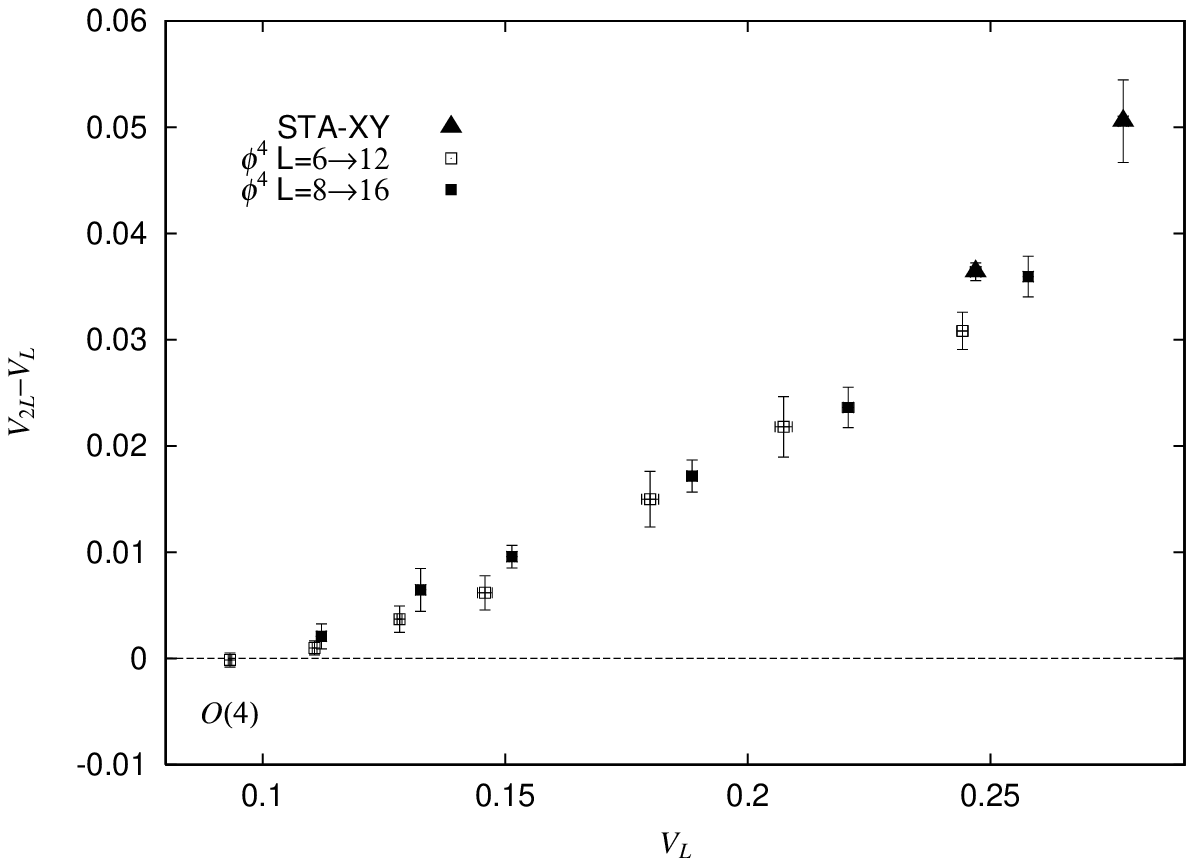}{12cm}{
The vertical velocity of the RG flow
$V_{2L}-V_L$ against $V_L$ for the
$N=2$ chiral $\phi^4$ model and the STA-XY model.
``$O(4)$'' denotes the $O(4)$--symmetric fixed point.
}

\myfig{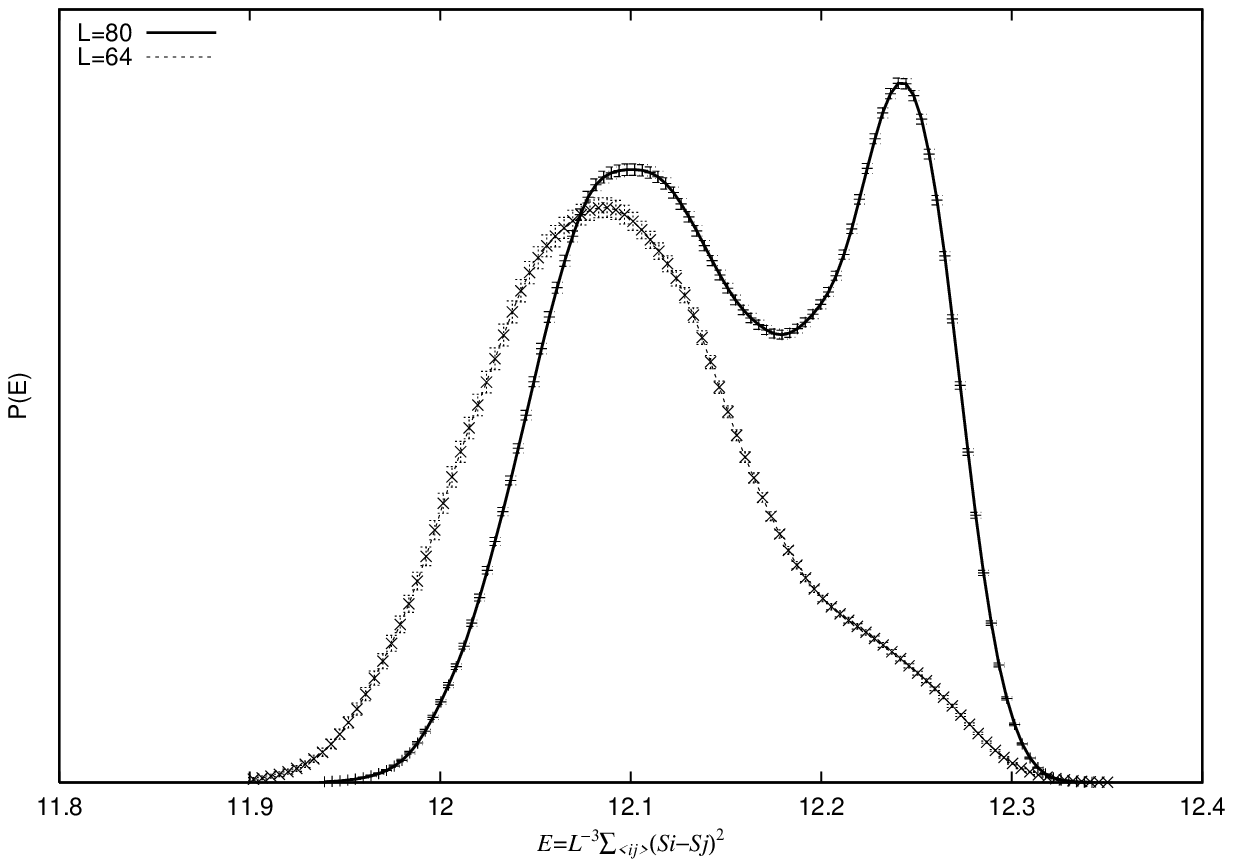}{12cm}
{
Histogram of
energy per spin of the $V_{3,2}$ model with the size
$L=64$ and $L=80$ at $K=0.217685$.
}

\myfig{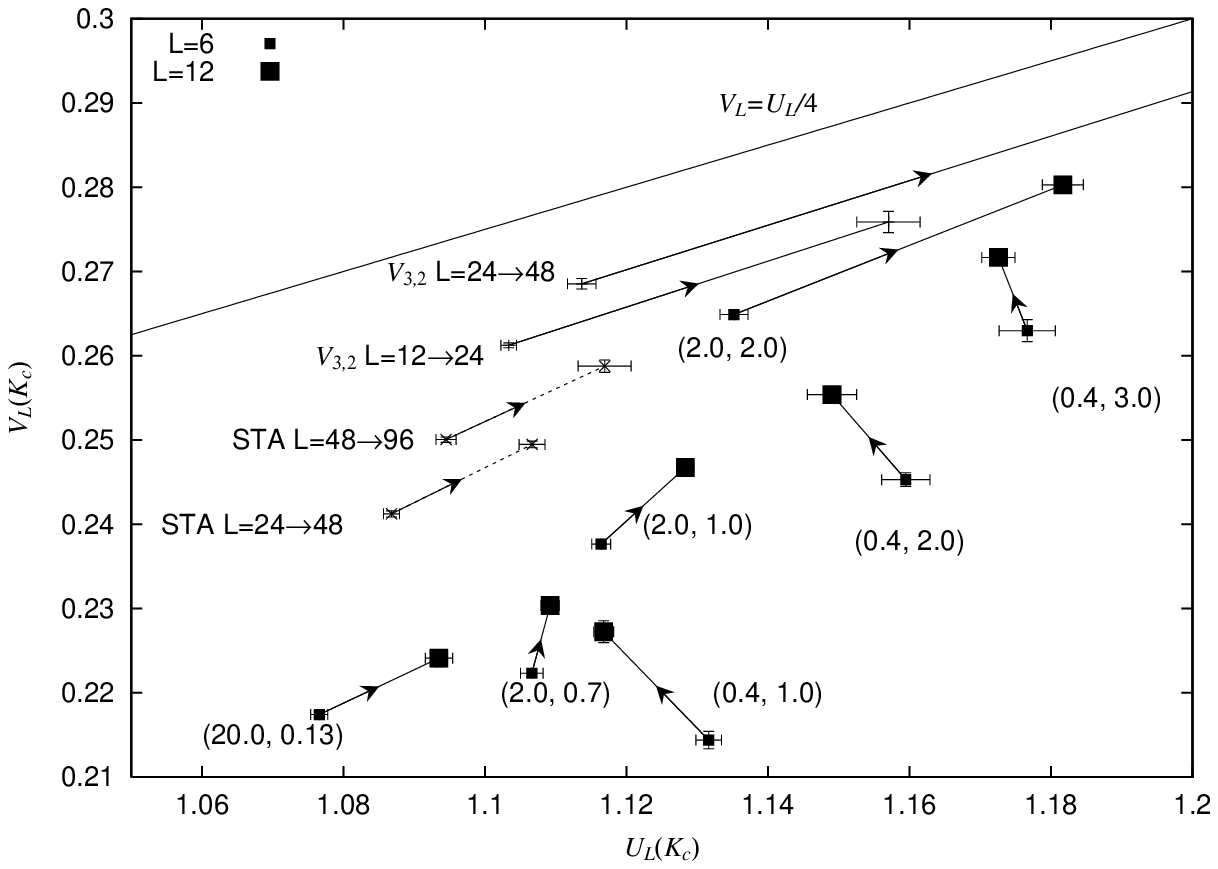}{10cm}
{
RG flow of 
the $N=3$ chiral $\phi^4$ model
and the STA-Heisenberg model
at the strong anisotropy region.
The numbers in parenthesis are the values of $(\lambda, A)$
of the chiral $\phi^4$ model.
}

\myfig{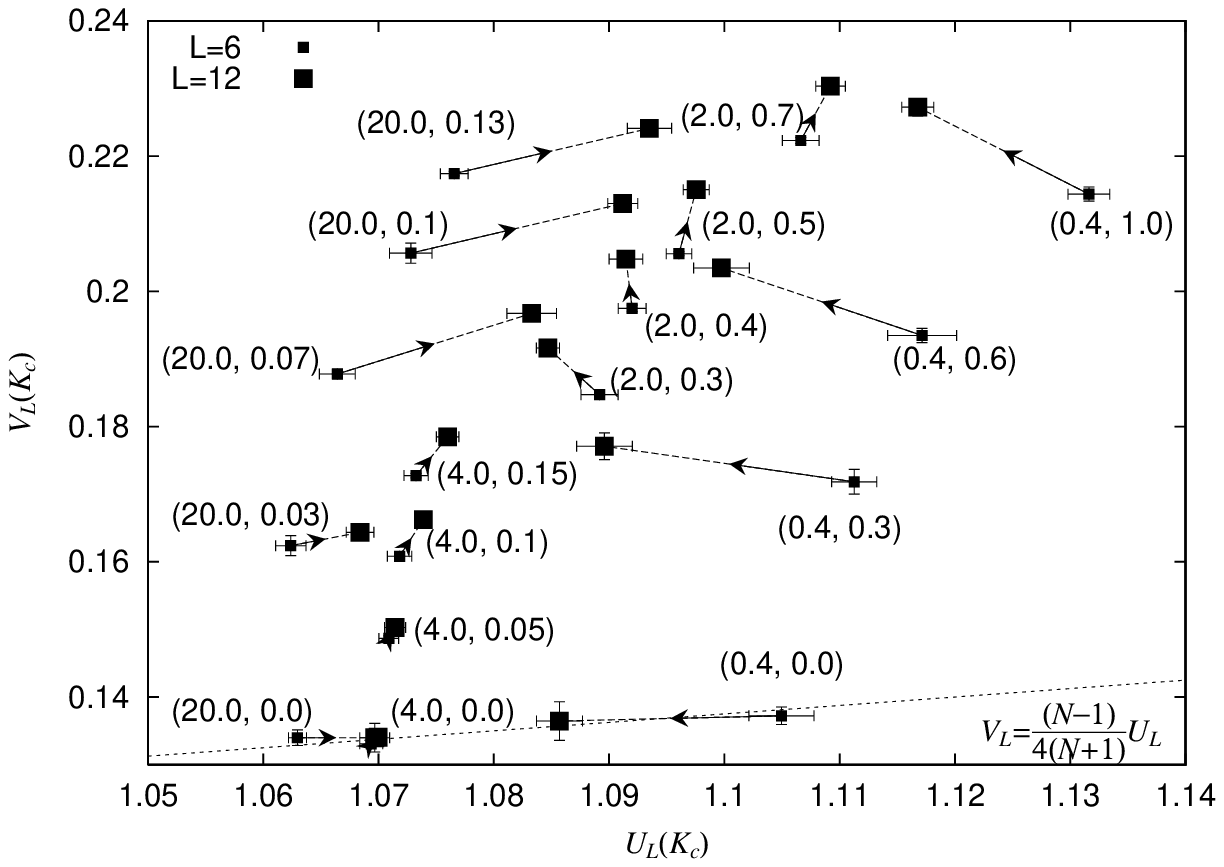}{10cm}
{
RG flow of 
the $N=3$ chiral $\phi^4$ model
at the weak anisotropy region.
The numbers in parenthesis are the values of $(\lambda, A)$.
}

\myfig{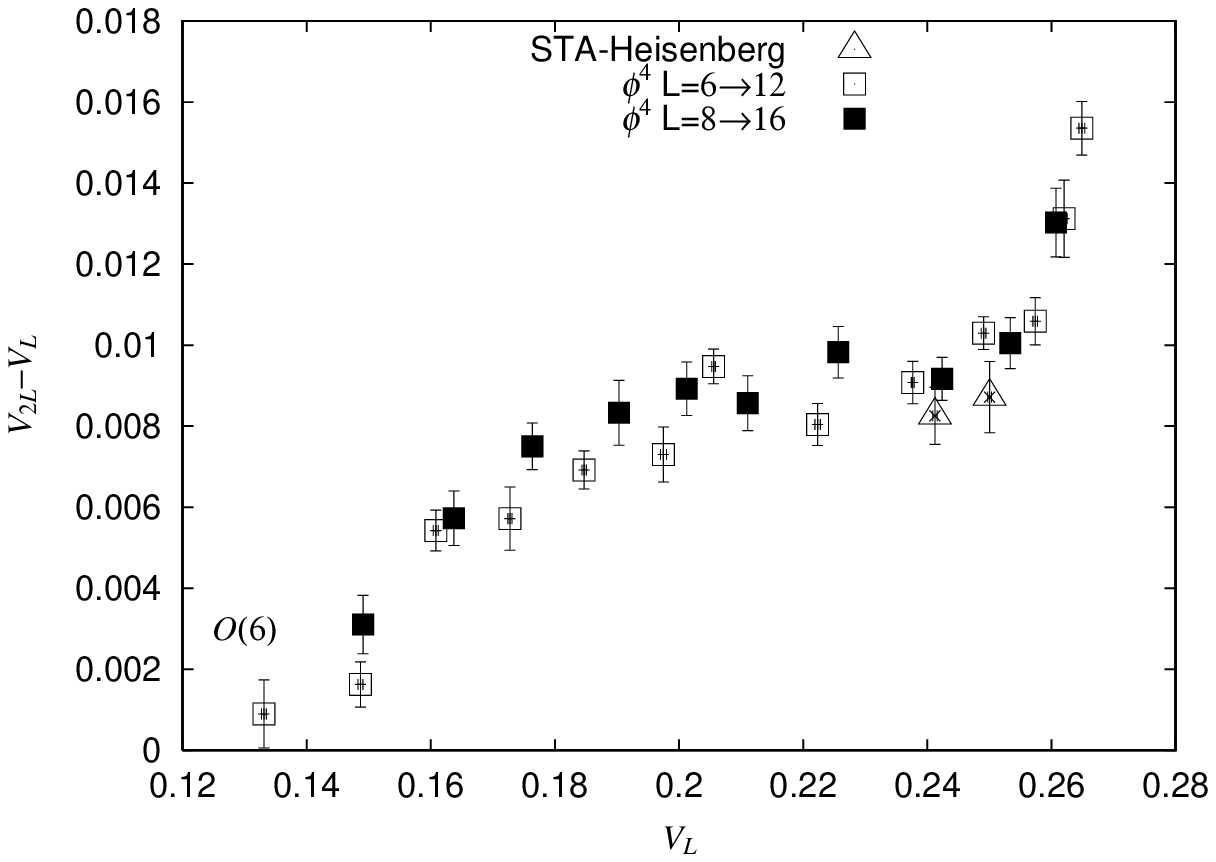}{12cm}{
The vertical velocity of the RG flow
$V_{2L}-V_L$ against $V_L$ for the
$N=3$ chiral $\phi^4$ model and the STA-Heisenberg model.
``$O(6)$'' denotes the $O(6)$--symmetric fixed point.
}

\myfig{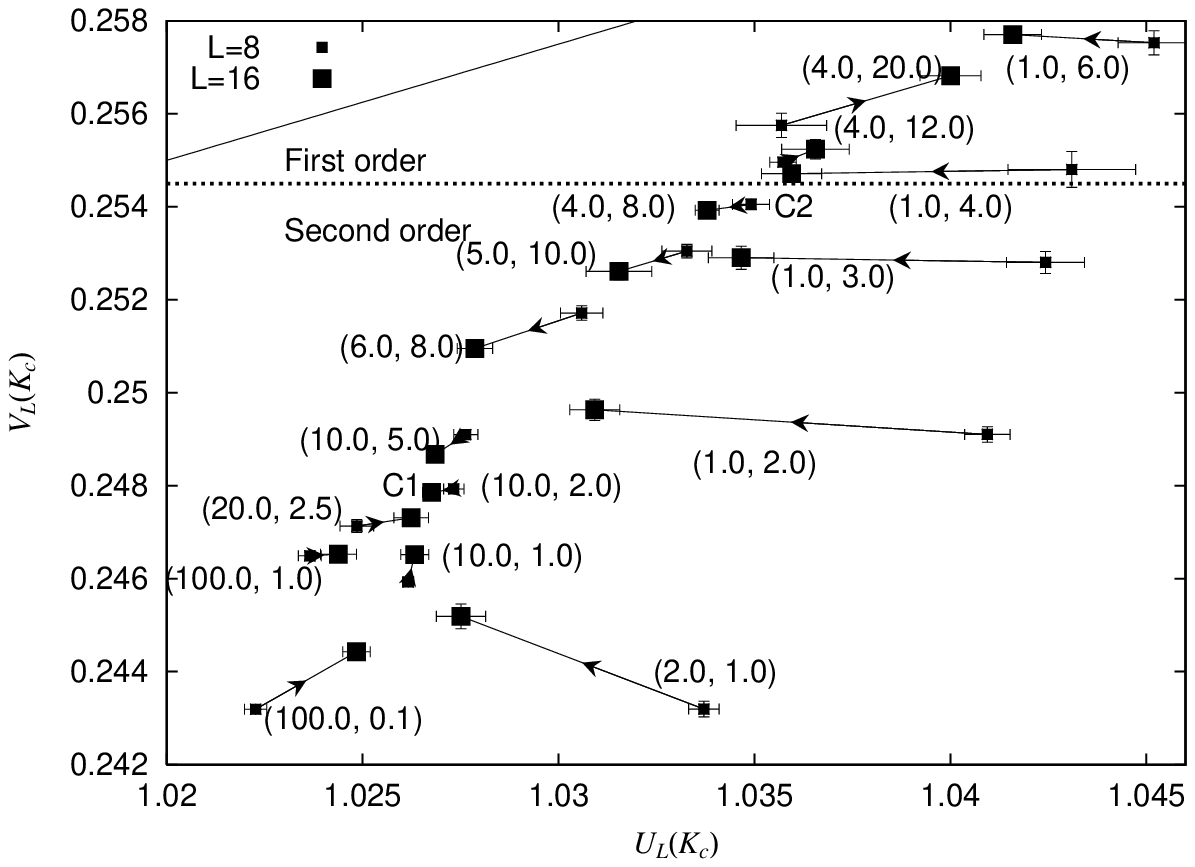}{10cm}
{
The RG flow of the $N=8$ chiral $\phi^4$ model.
The numbers in parenthesis show the values of $(\lambda, A)$.
C1 and C2 denote the stable and unstable RG fixed point,
respectively.
}

\myfig{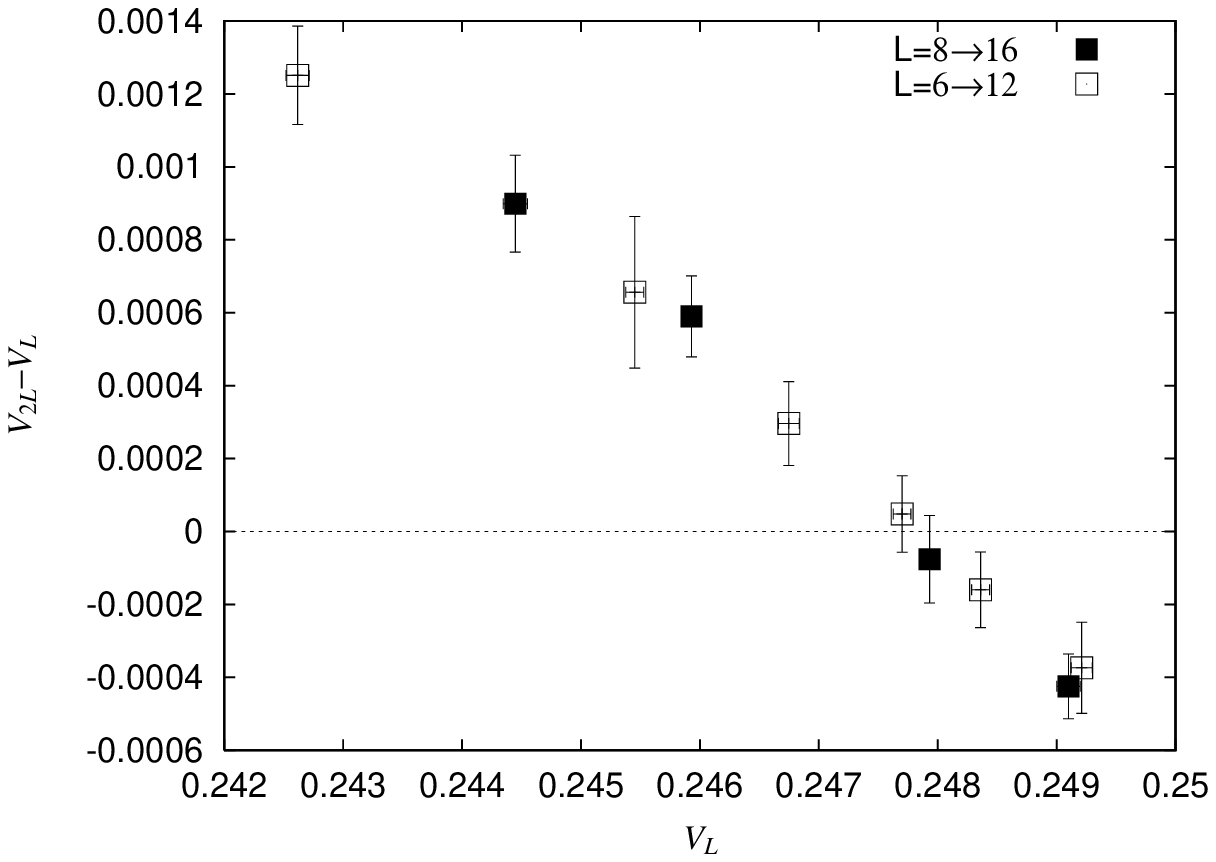}{10cm}{
The vertical velocity of the RG flow
$V_{2L}-V_L$ against $V_L$ near the stable chiral
fixed point $C_1$ of the $N=8$ chiral $\phi^4$ model.}

\newpage

\begin{table}[htbp]
\begin{center}
\begin{minipage}{7cm}
\begin{tabular}{|l|l|l|l|}
$\lambda$ & $A$ & $K/N$ & $C_{INT}$ \\ \hline
0.4 & 0.0 & 0.346 & 0.18\\
0.4 & 0.1 & 0.371 & 0.18\\
0.4 & 0.3 & 0.415 & 0.18\\
0.4 & 0.5 & 0.462 & 0.21\\
0.4 & 0.8 & 0.535 & 1.4\\
0.4 & 1.4 & 0.659 & 1.4\\
0.4 & 2.0 & 0.791 & 2.1\\
\hline
2.0 & 0.0 & 0.2880 & 0.12\\
2.0 & 0.1 & 0.3095 & 0.12\\
2.0 & 0.2 & 0.3292 & 0.12\\
2.0 & 0.3 & 0.3480 & 0.12\\
2.0 & 0.5 & 0.3800 & 0.32\\
2.0 & 0.7 & 0.4090 & 0.32\\
2.0 & 1.0 & 0.4510 & 0.40\\
\hline
20.0 & 0.00 & 0.2385 & 0.14\\
20.0 & 0.02 & 0.2415 & 0.14\\
20.0 & 0.06 & 0.2450 & 0.21\\
20.0 & 0.09 & 0.2460 & 0.35\\
\hline
\multicolumn{2}{|c|}{STA} & 0.7726 & 0.10
\end{tabular}
\end{minipage}
\end{center}
\caption{\label{n2param} 
Parameters $\lambda$, $A$, and $K$ (divided by $N$)
at which simulations were carried out 
for the $N=2$ chiral $\phi^4$ model
and the STA-Heisenberg model.
Physical quantities were observed for every
$C_{INT} \times L^2$ MCS.
}
\end{table}

\begin{table}[htbp]
\begin{center}
\begin{minipage}{7cm}
\begin{tabular}{|l|l|l|l|}
$\lambda$ & $A$ & $K/N$ & $C_{INT}$\\ \hline
0.4 & 0.00 & 0.3185 & 0.50 \\
0.4 & 0.30 & 0.3660 & 0.60 \\
0.4 & 0.60 & 0.4130 & 1.2 \\
0.4 & 1.00 & 0.4700 & 1.8 \\
0.4 & 2.00 & 0.6140 & 1.8 \\
0.4 & 3.00 & 0.7556 & 3.0 \\
\hline
2.0 & 0.3 & 0.3218 & 0.14\\
2.0 & 0.4 & 0.3328 & 0.21\\
2.0 & 0.5 & 0.3429 & 0.21\\
2.0 & 0.7 & 0.3624 & 0.35\\
2.0 & 1.0 & 0.3890 & 0.35\\
2.0 & 2.0 & 0.4640 & 1.5 \\
\hline
4.0 & 0.00 & 0.2650 & 0.14\\
4.0 & 0.05 & 0.2725 & 0.14\\
4.0 & 0.10 & 0.2776 & 0.14\\
4.0 & 0.15 & 0.2833 & 0.14\\
\hline
20.0 & 0.00 & 0.242 & 0.14\\
20.0 & 0.07 & 0.247 & 0.14\\
20.0 & 0.10 & 0.247 & 0.21\\
20.0 & 0.13 & 0.247 & 0.21\\
\hline
\multicolumn{2}{|c|}{$V_{3,2}$} & 0.2176 & 1.0\\
\hline
\multicolumn{2}{|c|}{STA} & 1.17286 & 0.05
\end{tabular}
\end{minipage}
\end{center}
\caption{\label{n3param} 
Parameters $\lambda$, $A$, and $K$ (divided by $N$)
at which simulations were
carried out 
for the $N=3$ chiral $\phi^4$ model
and the STA-Heisenberg model.
Physical quantities were observed for every
$C_{INT} \times L^2$ MCS.
}
\end{table}

\begin{table}[htbp]
\begin{center}
\begin{minipage}{7cm}
\begin{tabular}{|l|l|l|l|}
$\lambda$ & $A$ & $K/N$ & $C_{INT}$ \\ \hline
1.0 & 2.0 & 0.3710 & 0.6\\
1.0 & 3.0 & 0.4100 & 0.8\\
1.0 & 4.0 & 0.4505 & 1.6\\
1.0 & 6.0 & 0.5270 & 2.4\\
\hline
2.0 & 1.0 & 0.3135 & 0.4\\
\hline
4.0 &  5.0 & 0.3355 & 0.4\\
4.0 &  8.0 & 0.3475 & 0.8\\
4.0 & 12.0 & 0.3584 & 1.2\\
4.0 & 20.0 & 0.3710 & 2.0\\
\hline
5.0 & 10.0 & 0.324 & 0.8\\
\hline
6.0 & 5.0 & 0.3012 & 0.6\\
6.0 & 8.0 & 0.3044 & 0.7\\
\hline
10.0 & 0.6 & 0.2675 & 0.15\\
10.0 & 1.0 & 0.2690 & 0.15\\
10.0 & 1.5 & 0.2700 & 0.15\\
10.0 & 2.0 & 0.2710 & 0.4\\
10.0 & 3.0 & 0.2718 & 0.4\\
10.0 & 5.0 & 0.2728 & 0.6\\
\hline
20.0 & 2.5 & 0.2525 & 0.6\\
\hline
100.0 & 0.1 & 0.2416 & 0.2\\
100.0 & 1.0 & 0.2380 & 2.0 
\end{tabular}
\end{minipage}
\end{center}
\caption{\label{n8param} 
Parameters $\lambda$, $A$, and $K$ (divided by $N$)
at which simulations were carried out 
for the $N=8$ chiral $\phi^4$ model.
Physical quantities were observed for every
$C_{INT} \times L^2$ MCS.
}
\end{table}


\begin{thebibliography}{99}
%                   book  volume   year  page
\newcommand{\cit}[4]{{#1} {\bf #2}, #3 (#4).}
\newcommand{\citt}[4]{{#1} {\bf #2}, #3 (#4)}

\def\prl{Phys. Rev. Lett.}
\def\prb{Phys. Rev. B}
\def\pre{Phys. Rev. E}
\def\pr{Phys. Rev.}
\def\jpsj{J. Phys. Soc. Jpn.}
\def\epj{Euro. Phys. J.}
%%%%%%%%%%%%%%%%%%%%%%%%%%%%%%%%%%%%%%%%%%%%%%%%%%%%%%%%%%%%%%%%%%%%%
\bibitem{tissier2000}
M.~Tissier, B.~Delamotte, and D.~Mouhanna,
\prl {\bf 84}, 5208 (2000);
preprint, {\tt cond-mat/0107183}.%;{\tt cond-mat/0101167}.

\bibitem{pelissetto2001}
A.~Pelissetto, P.~Rossi  and E.~Vicari
\prb {\bf 63}  R140414 (2001); 
Nucl. Phys. B {\bf 607}, 605 (2001); 
\prb {\bf 65} 020403 (2002).
%preprint, {\tt cond-mat/0106525}.

%\bibitem{var2001}
%K.~B.~Varnashev, preprint.

\bibitem{plakhty2000}
V.~P.~Plakhty, J.~Kulda, D.~Visser, E.~V.~Moskvin, and J.~Wosnitza,
\prl {\bf 85}, 3942 (2000).

\bibitem{loison2000}
D.~Loison and K.~D.~Schotte,
\epj B {\bf 14}, 125 (2000).

\bibitem{kawamurareview}
H.~Kawamura,
J. Phys.: Condens. Matter {\bf 10}, 4707 (1998).

\bibitem{kawa-ft}
H.~Kawamura,
\jpsj {\bf 59}, 2305 (1990).

\bibitem{antonenko2}
S.~A.~Antonenko, A.~I.~Sokolov and K.~B.~Varnashev,
Phys. Lett. A {\bf 208}, 161 (1995).

\bibitem{antonenko1}
S.~A.~Antonenko and A.~I.~Sokolov,
\prb {\bf 49}, 15901 (1994).

\bibitem{zumbach1993}
G.~Zumbach, \prl {\bf 71}, 2421 (1993);
Nucl. Phys. B {\bf 413}, 771 (1994).

\bibitem{kawamuramc}
H.~Kawamura,
\jpsj {\bf 61}, 1299 (1992).

\bibitem{plumer}
M.~L.~Plumer and A.~Mailhot,
Phys. Rev. B {\bf 50}, 16113 (1994).

\bibitem{prefactor}
In the homogenious models with translational symmetry,
the prefactor $\langle \vec{\phi}({\bf x};l)^2 \rangle$ do not depend
on the position ${\bf x}$. In the inhomogenious models
such as random bond or diluted spin models, this factor
should be replaced by $[\langle \vec{\phi}({\bf x};l)^2 \rangle ]$
where $[\cdots ]$ denotes average over ${\bf x}$ and the randomness
in the Hamiltonian.

\bibitem{mcrg-conv}
R.H.Swendsen,
\cit{\prl}{42}{859}{1979}

\bibitem{mcrg}
M.~Itakura,
\pre {\bf 61}, 5924 (2000).

\bibitem{note1}
Alternatively, one can use this condition as a
definition of the fixed point.
The critical value of the relevant scaling fields
do not depends on the definition.

%BINDER
\bibitem{binder} K.~Binder,
\cit{Z. Phys. B}{43}{119}{1981}

\bibitem{q4ita} For example, see
M.~Itakura,
\cit{\prb}{60}{6558}{1999}


\bibitem{kaf}
The first Brillouin zone is an (approximately regular) hexagon
in the $x$--$y$ plane and $\vec{k}_{AF}$ is on the 
vertex of the hexagon.

\bibitem{ko01} K.~Kaneda and Y.~Okabe,
\cit{\prl}{86}{2134}{2001}

\bibitem{kok99} K.~Kaneda, Y.~Okabe, and M.~Kikuchi, 
 J. Phys. A {\bf 32}, 7263 (1999). 

\bibitem{wolff} U.~Wolff,
\cit{\prl}{62}{361}{1989}

\bibitem{caselle}
M.~Caselle and M.~Hasenbusch,
\cit{J. Phys. A}{31}{4603}{1998}

\bibitem{hist} A.~M.~Ferrenberg and R.~H.~Swendsen,
\cit{\prl}{61}{2635}{1988}

\bibitem{loisonxy} D.~Loison and K.~D.~Schotte,
\epj B{\bf 5}, 735 (1998).

\bibitem{plumer1d}
M.~L.~Plumer and  A.~Mailhot,
\cit{J. Phys. Condens. Matter}{9}{L165}{1997}

\bibitem{wang}
J.~Wang, D.~P.~Belanger, and B.~D.~Gaulin,
\cit{\prl}{66}{3195}{1991}

\end{thebibliography}
\end{document}